\documentclass[sigconf]{acmart}
\acmConference[EASE 2026]{The 30th International Conference on Evaluation and Assessment in Software Engineering}{9–12 June, 2026}{Glasgow, Scotland, United Kingdom}

\AtBeginDocument{%
  }

\usepackage{CommonPaper}

\usepackage{enumitem}

\usepackage{url}
\usepackage{graphicx}

\usepackage[T1]{fontenc}
\usepackage{amsmath}
\usepackage{newtxmath}
\usepackage{mathtools}

\usepackage{algorithm}
\usepackage{algpseudocode}

\usepackage[svgnames]{xcolor}

\usepackage{listings}
\definecolor{diffstart}{named}{Grey}
\definecolor{diffincl}{rgb}{0, 0.35, 0}
\definecolor{diffrem}{rgb}{0.72, 0, 0}
\lstdefinelanguage{diff}{
morecomment=[f][\color{diffstart}]{@@},
morecomment=[f][\color{diffincl}]{+},
morecomment=[f][\color{diffrem}]{-},
}

\usepackage{multirow}
\usepackage{makecell}

\usepackage{caption}
\usepackage{subcaption}

\usepackage{capt-of}

\newcommand*{\repo}[1]{{\it #1}}

\newcommand*{\Tool}[1]{\textsc{#1}}
\newcommand{\Summer}{\Tool{Summer}}
\newcommand*{\CE}[1]{\textcolor{gray}{\textbackslash #1}}
\def\magnet{\raisebox{-.1\height}{\includegraphics[height=0.7\baselineskip]{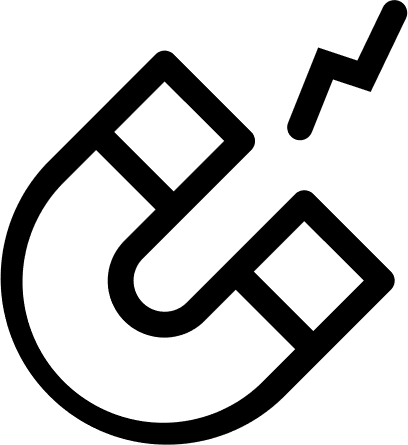}}}

\def\totalJava{106}

\def\nonJavaGradle{9}
\def\nonJavaMarkdown{16}
\def\nonJavaProperties{7}
\def\nonJavaXml{23}
\def\nonJavaLikeJava{5}
\def\nonJavaOthers{13}

\edef\totalNonJava{\fpeval{\nonJavaGradle+\nonJavaMarkdown+\nonJavaProperties+\nonJavaXml+\nonJavaLikeJava+\nonJavaOthers}}
\edef\datasetSize{\fpeval{\totalJava+\totalNonJava}}

\def\SummerMatchesJava{34}
\def\SummerMatchesNJ{30}
\edef\SummerAccuracyJava{\fpeval{round(\SummerMatchesJava / \totalJava,2)*100}\%}
\edef\SummerAccuracyNJ{\fpeval{round(\SummerMatchesNJ / \totalNonJava,2)*100}\%}
\edef\SummerAccuracyOverall{\fpeval{round((\SummerMatchesJava+\SummerMatchesNJ) / \datasetSize,2)*100}\%}

\newif\ifblind
\blindtrue

\title[A Universal Textual Merge Strategy Based on Tokens]{A Universal Textual Merge Strategy Based on Tokens\\ for Version Control Systems}

\author{Qiqi Jason Gu}
\email{qiqi.gu@cvut.cz}
\affiliation{
\institution{CIIRC, Czech Technical University in Prague}
\city{Prague}
\country{Czech Republic}
}

\author{Mikol\'a\v{s} Janota}
\email{mikolas.janota@cvut.cz}
\affiliation{
\institution{CIIRC, Czech Technical University in Prague}
\city{Prague}
\country{Czech Republic}
}

\begin{document}

\begin{abstract}
Merging is a core operation in version control systems such as Git,
but traditional line-based algorithms often yield spurious conflicts, particularly in the presence of refactorings or parallel edits.
While syntax- and semantics-aware merging approaches can reduce conflicts, they introduce drawbacks such as loss of formatting, dependence on language-specific parsers, and limited flexibility across heterogeneous artifacts.
To address this gap, we present {\Summer}, a novel textual token-based merge algorithm independent of document formats.
Dividing text into tokens,
our approach formulates token-level changes in one branch into string-rewriting rules and move rules,
and applies these rules to the text of the other branch to construct a merge.
Despite being independent on programming languages, our move rules model extracting and inlining functions.
We evaluated {\Summer} on ConflictBench, a large benchmark of real-world merge scenarios, comparing it with five pioneering merge tools across Java and non-Java files.
Experimental results show that {\Summer} achieved the highest {\SummerAccuracyOverall} accuracy in reproducing merges verbatim identical to developers',
and ranked second in semantic accuracy.
\end{abstract}




\maketitle

\section{Introduction}
Version control systems, such as Git and Perforce, are commonplace in
modern software development~\cite{seibt2021leveraging},
video game development,
and book writing~\cite{pe2018collaborative}.
They allow practitioners to work in parallel on separate branches and later integrate their work by performing a merge.
A successful merge is saved by a merge commit with two parent revisions.

Merging is a critical yet nontrivial operation in version control systems.
Apart from custom merge drivers, Git provides six built-in merge strategies, which all operate purely at the textual level.
Because changes are compared line by line,
if two parallel modifications are made in the same line, this line fails to merge and a conflict is reported~\cite{mens2002state}.
Beyond line-level edits, common programming practices such as code refactorings or method reordering further impede \Tool{git diff}'s ability in aligning lines,
and then given a misaligned diff, \Tool{git merge} produces conflicts when the intended changes are logically compatible~\cite{seibt2021leveraging}.

In spite of great advances in deep learning,
neural networks or large language models are not a panacea in merging revisions of repositories.
On the one hand, running neural networks locally requires high-end GPUs, which not every developer has.
On the other hand, sending data over to the cloud raises security and privacy concerns.
Partly due to these reasons,
current research mainly applies neural networks on small, local conflict zones~\cite{shen2023git,svyatkovskiy2022program,dinella2022deepmerge}
rather than merging two whole branches
where hundreds of files can be changed, along with file renaming and copying and modification to metadata.

While there is abundant literature on parsing and merging Java source code~\cite{ghiotto2018nature,accioly2018understanding,ellis2022operation},
support for non-Java languages remains limited due to inherent parsing difficulties.
For example, {\TeX} is context-sensitive, and determining its parse tree requires executing the code.
C++ is also context-sensitive.
For instance,
if \code{int f(M);} is preceded by \code{\#define M int}, then \code{f} is a function that takes an int and returns an int.
If it is preceded by \code{\#define M 0}, then \code{f} is an int variable initialized with 0.
Parsing dynamic languages, such as JavaScript, is also a challenge~\cite{dinella2022deepmerge}.

Next, even if a parse tree can be determined, it is common for a tool not to see the trees for the forest.
A smart, semantics-aware diff tool in C++ is sensitive to semantic differences but disregards syntactical disparities.
Given \code{value\CE{t}= 5[myArray]} and \code{value\CE{t}= *(myArray + 5)}
which are functionally equivalent because both express array indexing,
the tool reports no difference since syntactical information is discarded.
Similarly, a syntax-aware diff tool may ignore lexical disparities in
comments, white spaces, parentheses for grouping, and other so-called ``trivia''.
However, trivia is important for professional software engineers who want to review formatting changes and rewording in comments~\cite{miraldo2019efficient}.

Finally, an AST-based merge result hardly fully reflects the original manner of coding.
Because of the lost details during comparison,
it cannot respect the original style when writing back a merged syntax tree.
For the previous example, the semantic tree may be written out as \code{value = myArray[5]},
which matches neither of the two versions and misses the tab character (\CE{t}) on the left of the equal sign.

At the same time, an AST tool still resorts to a textual merge tool for low-level data,
including the content of string literal nodes,
and files that the tool does not understand,
including README files, configuration scripts, and various auxiliary artifacts, which are present in nearly all repositories~\cite{pan2021can}.
Hence, we propose a universal textual merge algorithm,
which can be used directly as a standalone merge driver by software engineers,
but also as a component called by an AST-based tool.
We name the algorithm and the implemented tool as {\Summer},
short for \textbf{Su}bstitution \textbf{M}anager and \textbf{M}erge \textbf{E}rror \textbf{R}esolver.

\begin{figure}
\centering
\includegraphics[width=0.6\linewidth]{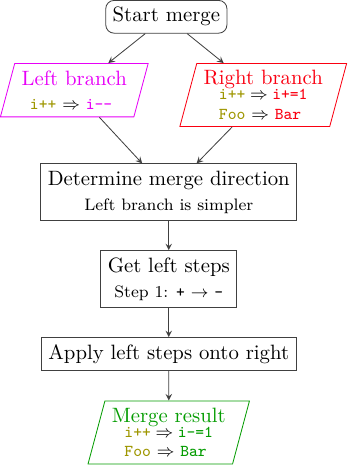}
\caption{The flowchart of the \code{merge} action in {\Summer}}
\label{fig:architecture}
\end{figure}

{\Summer} offers three user-level actions, \code{decompose}, \code{rebase}, and \code{merge}.
It can be installed as a plugin of Git, so that \code{git rebase} calls \code{summer rebase} and \code{git merge} calls \code{summer merge}.
\begin{description}
\item[decompose] It is the most fundamental action, which summarizes modification steps conducted in one or more commits.
For example, if a class \code{Foo} is renamed to \code{Bar}, this change may affect many files.
Nevertheless, a single step, \mbox{\code{Foo} $\rightarrow$ \code{Bar}}, is sufficient to reproduce the entire modification.
\item[rebase] This action invokes the action \code{decompose} to get modification steps and then applies these steps onto another branch.
\item[merge] Given two branches, the action \code{merge} first determines a merge direction (Section~\ref{sec:merge-direction}),
whether from left to right, or from right to left,
then invokes the action \code{rebase}.
\end{description}
Figure~\ref{fig:architecture} illustrates the workflow of merging the left branch onto the right branch.
Suppose a file in a base revision has content \code{i++} and \code{Foo}.
The left branch modifies \code{i++}  to \code{i-{}-}, denoted by $\Rightarrow$.
The right branch makes two changes, modifying \code{i++} to \code{i+=1} and \code{Foo} to \code{Bar}.
The process Determine merge direction
determines that the left branch contains only one change, making it simpler to decompose and apply.
Thus we decompose the changes in the left branch into steps.
The change \code{i++} $\Rightarrow$ \code{i-{}-} boils down to editing a single symbol, so the rule, denoted by  $\rightarrow$, is \code{+} $\rightarrow$ \code{-}.
Finally, the rule applies to the right branch and we get \code{i-=1} and \code{Bar} as the merge result.

Each step {\Summer} produces is
either a mathematical string-rewriting rule (Section~\ref{sec:simple-rewriting}), which rewrites the current revision towards the target revision,
or a move rule (Section~\ref{sec:move}), which says ``if a string-rewriting rule matches, then execute another string-rewriting rule.''

In Section~\ref{sec:conflictBench},
we employ a notable benchmark ConflictBench created by \etal{Shen}~\cite{shen2024conflictbench} for evaluating merge tools.
{\Summer} and 5 widely-used merge tools are compared on this benchmark.
The results show that our algorithm achieved the highest overall accuracy (\SummerAccuracyOverall) in Java and non-Java files when we evaluated character-to-character matching against developers' merge results.
If we compared {\Summer}'s literal match accuracy with language-dependent tools' semantic match accuracy,
the {\SummerAccuracyOverall} merge accuracy of {\Summer} is the second highest, with \Tool{AutoMerge} leading at 46.2\%.
Although individual developers emphasize semantic accuracy,
our tool's literal matching capability makes it ideal in an industrial setting.

We conclude that {\Summer} outperforms the majority of widely-used merge tools in terms of merge accuracy.
Section~\ref{sec:threats-to-validity} discusses threats to the validity of our findings.
The last section concludes our paper.

\section{Related Work}\label{sec:relatedWork}
Merging utilizes a common ancestor as the base and applies changes from one branch onto another.
In this section we study 4 aspects of a merge tool:
its mathematical theories,
scopes of merging,
how to use text,
and alignments of substrings or tree nodes.
We briefly discuss merge conflict resolutions, which are to try another merge strategy in a smaller scope.
The effectiveness of resolving a conflict locally may be limited due to an error in a broader scope,
such as a wrong pair of file names being aligned.

\subsection{Theories of Merging}

Darcs~\cite{roundy2005darcs} is a version control system
which stores patches (functions that transform text).
Darcs is built on top of patch theory.
Patch theory argues that the merge operation is commutative, i.e.,
merges from A to B and from B to A should yield the same result, either a conflict or the same textual state~\cite{jacobson2009formalization}.
The merge operation in Darcs has exponential running time~\cite{gentle2025collaborative}
because Darcs allows interactions among patches.
Mimram and Cinzia~\cite{mimram2013categorical} modeled merges as pushouts in category theory.
If the partial order of two added lines from two branches cannot be established, the two branches are in conflict.

Git is built on top of directed acyclic graphs, and stores static text for each revision.
The 3-way merge strategy in Git is commutative, but rebasing is not.
{\Summer} is designed for traditional version control systems such as Git and Subversion,
but {\Summer} presents Darcs's patches as string-rewriting rules, so that it can merge text in a powerful way.

\subsection{Scopes of Merging}
Although most merging tasks ultimately boil down to operations on plain text,
the scope of merging can vary considerably.
It ranges from a short fragment of text, a file identified by its path, an entire directory containing many files,
to a full revision with metadata in a version control repository.

As a wrapper around GNU diff, \Tool{wdiff} highlights differences word by word
by splitting each word in an input file into its own line, then runs GNU diff.
\Tool{wiggle}~\cite{wiggle} is the word-wise version of GNU patch, capable of incorporating two pieces of text with regard to a base.
These two tools work on the text level and do not compare or merge file names.

Folder-level tools include
\Tool{AutoMerge}~\cite{automerge},
\Tool{IntelliMerge}~\cite{IntelliMerge},
\Tool{FSTMerge}~\cite{FSTMerge1},
and \Tool{KDiff3}.
These tools compare two folders or merge three folders, treating one as the base.
However, none of them performs file renaming detection.

Stock merge tools in version control systems merge everything in a revision, including file renaming detection or tracking.
Git additionally has a ``rerere'' functionality that reuses historical merge resolutions.

\etal{Vale} advised merge tools to analyze an entire merge request rather than only the conflict parts
in order to get more context~\cite[p4979-4980]{vale2021challenges}.
Inspired by Programming by Example,
the algorithm by \etal{Pan}~\cite{pan2021can} learns merge patterns from the whole history of a repository
and applies those patterns to a particular merge scenario.

Resembling the stock merge tool in Git,
{\Summer} receives revision identifiers and merges everything in a revision.
It treats all data, whether file content or file names, as strings by a mapping (Section~\ref{sec:prepare}).

\subsection{Preprocessing of Text}

When a merging task boils down to plain text, the way to preprocess the text is another problem.
A piece of text can be broken into words, lines, or blocks.
With domain knowledge, it can also be parsed into a tree.

The most well-known \Tool{GIT/GNU diff} compares textual input line by line.
This approach has an obvious disadvantage in that it cannot handle parallel modifications to the same line~\cite{mens2002state}.
There is an early algorithm~\cite{tichy1984string}
that detects moved strings between the old and new text,
but it does not permit minor modifications within a moved string, such as renaming a variable.
\Tool{wdiff} and \Tool{wiggle}~\cite{wiggle} compare and merge words.

Parsing a piece of program code into a tree is extensively scoped.
\etal{Seibt}~\cite{seibt2021leveraging} showed that
compared to \Tool{git merge}, a syntactic merge strategy decreased the number of conflicts from 6.69\% to 5.32\%.
\Tool{CLDiff}~\cite{cldiff} parses Java code and builds an abstract syntax tree (AST).
Nodes at or higher than the statement level are merged with a tree algorithm;
other elements are passed to \code{git diff} and \code{git merge}.
Comments are ignored.
Employing a different cut-off,
\Tool{JDime}~\cite{jdime} incorporates method bodies in a textual way and incorporates high-level nodes syntactically,
but it has difficulty handling identifier renaming.
Furthermore, sometimes \Tool{JDime} performed a clean merge but the artifact turned out not compilable~\cite{seibt2021leveraging}.
\etal{Le{\ss}enich}~\cite{lessenich2017renaming} enhanced \Tool{JDime}'s handling of renaming and shifted code.
\Tool{FSTMerge} was initially proposed in \cite{FSTMerge1}, then improved in \cite{FSTMerge2}.
It can merge Java, C\#, and Python code and fall back to \Tool{git merge} for the remaining content.

From a semantic viewpoint, code refactoring leads to~22\% of merge conflicts~\cite{mahmoudi2019refactorings}.
\Tool{RefMerge}~\cite{ellis2022operation} and \Tool{IntelliMerge}~\cite{IntelliMerge}
are the state of the art for resolving conflicts of this particular kind.
They detect refactorings in two branches, undo them, then call \code{git merge} to merge the two branches,
and finally replay the refactorings on the merged codebase.

Although AST-based tools are blind to trivia such as comments and parentheses,
it is rare for merge tools to use concrete syntax trees (CST).
We only found one preprint~\cite{duarte2025lastmerge}, which is not yet peer-reviewed, that uses a CST for merging.
When designing a differencing algorithm for a language for cyber-physical systems,
Sj{\"o}lund~\cite{sjolund2021evaluating} weighed up a CST but gave it up due to the increased complexity.
Bates~\cite{bates2002text} built a text editor that parses the content into a CST
and only sends the modified part of the CST for updating the user interface.
Nevertheless, Bates' algorithm is not for a version control system.

{\Summer} fills the gap by dividing text into tokens of letters, digits, white spaces, and symbols before merging.

\subsection{Aligning Similarities}

To find differences, one must match unchanged or similar parts so that other parts are labeled as added or deleted.
For instance, \code{git diff} has options \code{-{}-find-renames}, \code{-{}-find-copies}, and \code{-{}-find-copies-\linebreak[1]harder} to perform alignments on the file level.

Myers' differencing algorithm~\cite{myers1986nd} aims for the shortest edit script.
Much as the number of additions or deletions is minimized, the content of each addition or deletion can be long.
Git implements multiple variations of Myers algorithm,
and one of them is the histogram diff algorithm, which according to \etal{Nugroho}~\cite{nugroho2020different}, provides fast speed and good code alignment.
{\Summer} first utilizes the histogram algorithm to obtain a line-wise diff,
then tokenizes changed lines, and calls the Levenshtein distance to get a token-wise diff (Section~\ref{sec:dissect}).

The problem of edit distance for unordered trees is NP-hard~\cite{zhang1994some}.
Consequently, researchers who study syntax tree-based merging
have to make trade-offs or heuristics to achieve accuracy and runtime speed.
If the bi-gram similarity~\cite{BigramSimilarity} of two nodes is greater than 0.8,
\Tool{CLDiff}~\cite{cldiff} aligns the two subtrees.
\Tool{IntelliMerge}~\cite{IntelliMerge} only aligns nodes of the same syntactic type.
\Tool{IntelliMerge} calculates the string cosine similarity for identifier nodes and the Jaccard similarity for other nodes,
with the matching threshold 0.618.
\etal{Zhu}~\cite{automerge} built \Tool{AutoMerge} on top of \Tool{JDime}, and matched nodes based on an adjustable quality function.

\etal{Dinella} invented \Tool{DeepMerge}~\cite{dinella2022deepmerge} which uses a neural network to align JavaScript lines in order to resolve a conflict in a local scope.
\Tool{DeepMerge} achieved 61\% accuracy if a conflict zone had fewer than 7 lines, and otherwise its overall accuracy was 36.5\%.
Nevertheless, their neural network only repositions existing lines rather than generating new code.

\section{The Workflow of Merge}\label{sec:algorithm}

The merge action of {\Summer} executes in 6 stages, corresponding to 6 subsections here.
An overview is available in Figure~\ref{fig:architecture} in Introduction.
Section~\ref{sec:prepare} covers preparation.
Section~\ref{sec:merge-direction} is to determine a merge direction.
The Get step process in Figure~\ref{fig:architecture} maps to Section~\ref{sec:dissect} to \ref{sec:move}.
Finally we talk about applying steps in Section~\ref{sec:apply-rules}.

{\Summer} accepts 3 revision identifiers (the SHA of a commit in Git)
for the left branch, the right branch, and the base of the two branches.
Since it is always possible to squash multiple commits into one,
in this section we describe our algorithm in terms of 3 commits, abandoning the term ``branch''.
We use the term ``commit'' to emphasize the modification, and the term ``revision'' to emphasize the state after the modification.
Our implementation supports unsquashed commits.

\subsection{Mapping a Commit to String Edits}\label{sec:prepare}

When asked to decompose one commit,
{\Summer} internally invokes its decomposition library.
Although {\Summer} handles file contents, file names, and metadata of a version control system,
the decomposition library works purely on strings. Therefore, a mapping from a commit to strings is required.

\begin{table}
\caption{A dictionary presenting a Git commit, where a submodule path is changed, a file is renamed, and its content is changed}
\label{table:keyValuePairs}
\begin{tabular}{p{0.24\linewidth}|p{5cm}}
\textbf{Label}             & \textbf{Bucket} \\\hline\hline
Submodule path           & github.com/txaty/bigcomplex $\Rightarrow$\newline gitlab.com/txaty/bigcomplex \\\hline
File name                & bc.go $\Rightarrow$ Program.go      \\\hline
Content of\newline bc.go &
\begin{lstlisting}[language=go, frame=tblr, aboveskip=-10pt, belowskip=0pt]
import bc "github.com/txaty/bigcomplex"

func main() {
	g1 := bc.NewGaussianInt(5, 6) // 5 + 6i
	g2 := bc.NewGaussianInt(1, 2) // 1 + 2i
	div := new(bc.GaussianInt).Div(g2, g1)
	fmt.Println(div)
}
\end{lstlisting}
$\Rightarrow$
\begin{lstlisting}[language=go, belowskip=-10pt, frame=tblr]
import bc "gitlab.com/txaty/bigcomplex"
import "fmt"

func main() {
	g1 := bc.NewGaussianInt(5, 6) // 5 + 6i
	g2 := bc.NewGaussianInt(1, 2) // 1 + 2i
	res := new(bc.GaussianInt).Div(g2, g1)
	fmt.Println(res)
}
\end{lstlisting}
\end{tabular}
\end{table}

Suppose in a Git commit, a submodule path is changed from \code{github.com/txaty/bigcomplex} to \code{gitlab.com/txaty/bigcomplex},
a file is renamed from \code{bc.go} to \code{Program.go},
and the file content is modified from \code{import bc "github.com...Println(div)\CE{n}\}} to \code{import bc "gitlab.com...Println(res)\CE{n}\}}.
{\Summer} reads the commit and formulates a dictionary as Table~\ref{table:keyValuePairs} shows.
The column Bucket contains the set of modifications in the string format;
the column Label records the meaning of each bucket.
Only the Bucket part is sent to the decomposition library to get fine-grained steps,
and the library thus processes strings in an agnostic way, whether they are file names or file contents.

\subsection{Determining Merge Direction}\label{sec:merge-direction}

Given a base commit, a left commit, and a right commit, {\Summer} has to determine whether to merge the left commit onto the right, or merge the right commit onto the left.
For example, if one commit deletes a file, and another commit modifies the same file, {\Summer} has to merge the former on top of the latter
as a modification cannot be applied to a non-existent file.
Similarly, if one commit deletes file A and modifies file B, and another commit modifies file A and deletes file B,
then neither direction will work and thus a conflict is raised.

If there is no file deletion in both commits, {\Summer} calculates the Levenshtein Distance
from the left to the base,
and from the right to the base.
Afterward, {\Summer} chooses to decompose the simple commit, and applies the steps onto a complicated commit.

\subsection{Dissecting String Edits}\label{sec:dissect}
The decomposition library receives a set of buckets and each bucket currently contains one single string edit.
Although each string edit can be directly returned as a step for merging, it is too coarse-grained.
For instance,  \code{import bc "github.com...Println(div)\CE{n}\}} $\Rightarrow$ \code{import bc "gitlab.com...\linebreak[1]Println(res)\CE{n}\}}
captures the context of a whole file, which is even worse than a traditional line-wise diff.
Therefore, we have to dissect the string edit in each bucket.

For each string edit, we call Git's histogram algorithm to get a diff.
According to \etal{Nugroho}~\cite{nugroho2020different}, the histogram algorithm is fast and provides good code alignment.

Next, for each block of consecutive deleted lines followed by consecutive added lines,
we tokenize the deleted and added lines.
We define four categories of characters: digit, letter, white space, and symbol.
Continuous characters of the same category form a token, except that each symbol forms its own token.
For example, \code{n=0xFF\_0f} is tokenized to
$\{$\code{n},
\code{=},
\code{0},
\code{xFF},
\code{\_},
\code{0},
\code{f}$\}$.

Finally, the two sequences of tokens from a modified block are sent to the Levenshtein distance algorithm for alignment,
so that we get a list of string edits about inserted, modified, deleted, and unchanged tokens for each bucket.

Displaying the three buckets of string edits,
Table~\ref{table:dissect} is the result of dissecting the column Bucket of Table~\ref{table:keyValuePairs}.
We retain identity string edits, i.e., the lhs and the rhs are identical,
but we visually write them in a smaller font size.
$\varepsilon \Rightarrow$ \code{import "fmt"\CE{n}} is an insertion edit that changes the empty string to an import statement.
Others are substitution edits.

\begin{table}
\newcommand{\Keep}[1]{\textcolor{darkgray}{\small #1\ $\Rightarrow\ $#1}}
\caption{Breaking up coarse-grained string edits from Table~\ref{table:keyValuePairs} into substitution edits, identity edits, and one insertion edit}
\label{table:dissect}
\begin{tabular}{p{6.5cm}}
\textbf{Buckets of string edits} \\\hline\hline
\texttt{github} $\Rightarrow$ \texttt{gitlab}\newline
\Keep{.com/txaty/bigcomplex}
\\\hline
\texttt{bc} $\Rightarrow$ \texttt{Program}\newline
\Keep{.go}
\\\hline
\Keep{ import bc "}\newline
\texttt{github} $\Rightarrow$ \texttt{gitlab}\newline
\Keep{ .com/txaty/bigcomplex"\CE{n}}\newline
$ \varepsilon \Rightarrow$ \texttt{import "fmt"\CE{n}}\newline
\Keep{ func main()// 1 + 2i\CE{n}\CE{t}}\newline
\texttt{div} $\Rightarrow$ \texttt{res}\newline
\Keep{ := new...\CE{n}\CE{t}fmt.Println(}\newline
\texttt{div} $\Rightarrow$ \texttt{res}\newline
\Keep{)\CE{n}\}}
\end{tabular}
\end{table}

\subsection{From String-Rewriting Instances to String-Rewriting Rules}\label{sec:simple-rewriting}

The string edits that we have shown, such as
\code{bc.go} $\Rightarrow$ \code{Program.go}
and
\code{)\CE{n}\}}  $\Rightarrow$ \code{)\CE{n}\}}, are mathematically called string-rewriting instances.
They describe \textit{what changed} but not a general pattern that can be applied elsewhere.
To merge branches, we need to transform these specific edits into reusable transformation rules,
which we call string-rewriting rules.
We use $\Rightarrow$ to denote a string-rewriting instance and $\rightarrow$ to denote a string-rewriting rule.
We call them rewriting instances and rewriting rules for short.

The previous section has dissected and aligned string-rewriting instances,
which creates a solid foundation to execute Algorithm~\ref{algo:substitution}, reversing each rewriting instance back to a sequence of rewriting rules.

\begin{algorithm}[t]
\begin{algorithmic}
\Function{getPreciseRewriting}{$\mathit{buckets}$}
\State  $P \gets \varnothing $
\ForAll{$b \in \mathit{buckets}$}
\For{$ i=0, \dots, |b|$}
\State \Call{expandEdit}{$i, b, \mathit{buckets}, P$}
\EndFor
\EndFor
\State \Return \Call{SortAndFilter}{$P$}
\EndFunction
\Function{expandEdit}{i, b, buckets, P\_ref}
\For{$j=0, \dots, \mathit{w}$}
\For{$k=0, \dots, \mathit{w}$}
\State $r \gets \Bigl(\operatorname{join}_l(\text{b[i-j:i+k]}) \to  \operatorname{join}_r(\text{b[i-j:i+k]}) \Bigl)$
\State $\mathit{tp},\mathit{fp} \gets$ \Call{getClassificationMetrics}{r, buckets}
\State P\_ref[r]=($\mathit{tp},\mathit{fp}$)
\EndFor
\EndFor
\EndFunction
\end{algorithmic}
\caption{find a list of precise rewriting rules from buckets of rewriting instances}
\label{algo:substitution}
\end{algorithm}

The function \Call{getPreciseRewriting}{} in Algorithm~\ref{algo:substitution} returns a set of rewriting rules from a set of bucketed string edits.
For each edit in each bucket, we call the function \Call{expandEdit}{} to reverse it to zero or more string-rewriting rules,
and check true positives ($\mathit{tp}$) and false positives ($\mathit{fp}$) of these rules.
In the function \Call{expandEdit}{},  $\mathit{w}$ is short for context window. The function $\operatorname{join}()$ takes each non-identity rewriting instance as a whole or takes individual tokens from identity rewriting instances.
$\operatorname{join}_l()$ joins the left-hand sides of rewriting instances in the specified range to a single string,
and $\operatorname{join}_r()$ joins right-hand sides.
Then, the $\rightarrow$ connective makes the two strings into one string-rewriting rule which is assigned to $r$.

Following Table~\ref{table:dissect},
\code{github} $\Rightarrow$ \code{gitlab} is passed into \Call{expandEdit}{}().
When $j=0$ and $k=0$, a rewriting rule \code{github} $\rightarrow$ \code{gitlab} is formed with $\mathit{tp}=2$ and $\mathit{fp}=0$.

\code{bc} $\Rightarrow$ \code{Program} is also passed into \Call{expandEdit}{}().
When $j=0$ and $k=0$, a rewriting rule \code{bc} $\rightarrow$ \code{Program} is formed.
Unfortunately, this rule incorrectly rewrites \code{import bc "github.com/txaty/bigcomplex"} to \code{import Program "github.com/\linebreak[1]txaty/bigcomplex"},
which boosts false positives.
When $k=1$, $\operatorname{join}_l()$ and $\operatorname{join}_r()$ add additional context from what comes after.
After \code{bc} $\Rightarrow$ \code{Program} is \code{.go} $\Rightarrow$ \code{.go} , which is an identity rewriting instance.
Therefore, $\operatorname{join}()$ takes one token at a time, which is the dot (\code{.}),
and a new rewriting rule \code{bc.} $\rightarrow$ \code{Program.} is formed.
This rule has $\mathit{tp}=1, \mathit{fp}=0$.

Classification metrics cannot be computed for an insertion rule $ \varepsilon \rightarrow$ \texttt{import "fmt"\CE{n}}.
When $j=1, k=0$, the created substitution rule is \code{\CE{n}} $ \rightarrow$ \code{\CE{n}import "fmt"\CE{n}}
with $\mathit{tp}=1, \mathit{fp}=6$.
When $j=0, k=1$,  the created substitution rule is \code{func} $ \rightarrow$ \code{import "fmt"\CE{n}func}
with $\mathit{tp}=1, \mathit{fp}=0$.

The function \Call{SortAndFilter}{} calculates the precision of each rule, namely $\mathit{tp}/(\mathit{tp}+\mathit{fp})$.
Rules with precision no more than 0.5 are filtered out.
The remaining non-overlapping rewriting rules are the ones that best reproduce the target modifications that {\Summer} passed to the decomposition library.

If \Call{getPreciseRewriting}{}() returns rules with precision less than 1, these rules make mistakes amid rewriting.
\Call{getPreciseRewriting}{}() has to be called one more time to generate another set of rules to fix these mistakes.
In the end, if we still cannot fully reproduce the target modifications,
we increase the context window $\mathit{w}$.
All in all, it is always possible to rewrite a string (or a set of strings) to a target string (or a target set) by a sequence of string-rewriting rules.

In terms of the time complexity,
let $M$ be the total number of string edits in $\mathit{buckets}$
and $N$ be the total number of tokens in the lhs of $\mathit{buckets}$.
The average length of an edit is $\frac{N}{M}$.
\Call{getClassificationMetrics}{}() invokes a string-searching algorithm,
resulting in the time complexity $\frac{N}{M}N$.
We loop $\mathit{w}^2$ times in \Call{expandEdit}{}() and $M$ times in \Call{getPreciseRewriting}{}.
Therefore, the time complexity of the function \Call{getPreciseRewriting}{} in Algorithm~\ref{algo:substitution} is $O(\mathit{w}^2N^2)$,
irrelevant to the number of edits\footnote{We assume comparing two tokens is an elementary operation, but in reality the average number of characters in a token matters. Copying characters in $\operatorname{join}_l()$ and $\operatorname{join}_r()$ are ignored too.}.

\subsection{The Move Rules}\label{sec:move}

Although it is guaranteed to find a list of substitution rewriting rules to completely represent a set of string edits,
these rules do not capture the nuance of moving text.
In contrast, moving text is quite common in software development, in the form of extracting a method and inlining a method.

\begin{lstlisting}[language=diff, float, label={code:move}, caption={Left commit: two lines are extracted into a new function}]
 ...
   public void addListener(O obj) {
-    notNull(obj);
-    validate(obj);
+    runCheck(obj);
     Listeners.add(obj.getListener());
   }

+
+  public void runCheck(O obj) {
+    notNull(obj);
+    validate(obj);
+  }
 }
\end{lstlisting}

Consider an example presented in the \Tool{RefMerge} paper~\cite{ellis2022operation}:
the left commit extracts two lines into a new function \code{runCheck} as Listing~\ref{code:move};
the right commit changes the calling style \code{notNull(obj)} $\Rightarrow$ \code{obj.notNull()} and
\code{validate(obj)} $\Rightarrow$ \code{obj.validate()}.
The algorithm presented in Section~\ref{sec:simple-rewriting} finds three fine-grained steps from Listing~\ref{code:move},
shown in Figure~\ref{steps-without-move}.

\begin{figure}
\setlength{\belowdisplayskip}{0pt}
\setlength{\belowdisplayshortskip}{0pt}
\begin{align*}
\normalsize\text{Step 1:}&\; \text{add a new definition:} \\
&\;\small \text{\tt runCheck(O obj)\{\boxed{\tt notNull(obj); validate(obj);}\}} \\
\normalsize\text{Step 2:}&\;\small \text{\tt\boxed{\tt validate(obj);}\CE{n}\CE{t}\CE{t}Listeners} \rightarrow \\
&\; \text{\code{runCheck(obj);\CE{n}\CE{t}\CE{t}Listeners}} \\
\normalsize\text{Step 3:}&\;\small \text{\tt \CE{t}\CE{t}\boxed{\tt notNull(obj);}\CE{n}\CE{t}\CE{t}runCheck} \rightarrow \\
&\; \text{\tt \CE{t}\CE{t}runCheck}
\end{align*}
\caption{Decomposed steps without the move semantics from the left commit (Listing~\ref{code:move}). Boxed parts do not apply on the right revision.}
\label{steps-without-move}
\end{figure}

None of the 3 steps in Figure~\ref{steps-without-move} are correct with respect to the right revision.
Step 1 can match in the right revision, but it writes out the old calling style.
Step 2 and Step 3 do not even match because the calling style has been changed from \code{validate(obj)} to \code{obj.validate()}
and \code{notNull(obj)} to \code{obj.notNull()}.

One way to solve this problem is a string-rewriting algebra in Darcs Patch Theory which allows a string-rewriting rule to be rewritten by another rule.
However, we take another approach in this paper. In this section, we present move rules, which also solve the problem of moving text.

A move rule is in the form $a\vdash c$, read as ``if $a$ matches, then execute $c$''. Both $a$ and $c$ are string-rewriting rules.
Unlike rewriting rules discussed in the previous section,
the lhs of $a$ includes a capture group {\magnet} that can capture text amid matching,
and the rhs of $c$ has a backreference to this captured text.

\begin{figure}
\setlength{\belowdisplayskip}{0pt}
\setlength{\belowdisplayshortskip}{0pt}
\begin{align*}
\text{Step 1$'$:} &\; \text{move whatever before \code{Listeners}} \\
				& a=\text{\code{\{\CE{n}\CE{t}\CE{t} \magnet\ \CE{n}\CE{t}\CE{t}Listeners}} \rightarrow \\
				& \qquad \text{\code{\{\CE{n}\CE{t}\CE{t}runCheck(obj);\CE{n}\CE{t}\CE{t}Listeners}} \\
				& c=\text{\code{\CE{n}\}}} \rightarrow \text{\code{\CE{n}\CE{n} \magnet\ \CE{n}\}}} \\
\text{Step 2$'$:}  &\; \text{add the function header} \\
				& \text{\code{\CE{t}\}\CE{n}}} \rightarrow \\
				& \text{\code{\CE{t}\}\CE{n}\CE{n}\CE{t}public void runCheck(O obj) \{\CE{n}\CE{t}\CE{t}}} \\
\text{Step 3$'$:}  &\; \text{add the function footer} \\
				& \text{\code{\CE{n}\}}} \rightarrow \text{\code{\CE{n}\CE{t}\}\CE{n}\}}}
\end{align*}
\caption{Decomposed steps with the move semantics from the left commit (Listing~\ref{code:move}). They can apply on the right revision.}
\label{steps-with-move}
\end{figure}

With the new move semantics, {\Summer} finds one move step and two string-rewriting steps, shown in Figure~\ref{steps-with-move}.
Step 1$'$ is a move rule. The execution result of Step 1$'$ is Listing~\ref{code:after-move}.
The lhs of $a$ in the left revision
captures the function body \code{notNull(obj);\CE{n}\CE{t}\CE{t}\linebreak[1]validate(\linebreak[1]obj);}
and its rhs is similar to the rhs of the original Step 2.
The rhs of $c$ moves the captured function body to Line 6 and 7.

When the move step executes in the left revision, it captures {\magnet} and moves \code{notNull(obj);\CE{n}\CE{t}\CE{t}validate(obj);}.
When it executes in the right revision, it captures {\magnet} and moves \code{obj.notNull();\CE{n}\CE{t}\CE{t}\linebreak[1]obj.\linebreak[1]validate();}.
In this way, move rules adapt the local style in the current revision.
Accordingly, move rules should be created and executed before string-rewriting rules.

So far, the call site (Line 2 in Listing~\ref{code:after-move}) has been perfectly reconstructed, but the function body at Line 6 and 7 is bare.
Step 2$'$ adds the function header \code{public void runCheck(O obj)},
and Step 3$'$ adds the footer of a single indented curly bracket.
Overall, these rules generated from the left commit correctly apply to the right revision.

\begin{lstlisting}[language={}, float, label={code:after-move}, caption={The content after the move rule Step 1$'$ in the right revision},
numbers=left, xleftmargin=1.8em]
	public void addListener(O obj) {
		runCheck(obj);
		Listeners.add(obj.getListener());
	}

obj.notNull();
		obj.validate();
}
\end{lstlisting}

Despite being similar to an operation of one cut and one paste,
a move operation can handle one-to-many and many-to-one edits.
One-to-many means that one piece of text is deleted and its substring is added to many places.
In other words, $a$ is matched once, and $c$ is matched and executed one or more times.
One-to-many models inlining operations.
On the other hand, many-to-one models extraction operations.
Algorithm~\ref{algo:move} illustrates the idea.
\Call{getPreciseMove}{} is akin to \Call{getPreciseRewriting}{}.
In the loop body, \Call{findExtract}{} and \Call{findInline}{} are terminology borrowed from code refactoring,
although {\Summer} itself has no concept of refactoring.

\begin{algorithm}[t]
\begin{algorithmic}
\Function{getPreciseMove}{$\mathit{buckets}$}
\State $P \gets \varnothing $
\ForAll{$b \in \mathit{buckets}$ }
\For{$ i=0, \dots, |b|$}
\If{b[i] is insertion}
\State \Call{findExtract}{$i, b, \mathit{buckets}, P$}
\ElsIf{b[i] is deletion}
\State \Call{findInline}{$i, b, \mathit{buckets}, P$}
\EndIf
\EndFor
\EndFor
\State \Return \Call{SortAndFilter}{$P$}
\EndFunction
\end{algorithmic}
\caption{Generate move rules from string edits organized in buckets}
\label{algo:move}
\end{algorithm}

Algorithm~\ref{algo:findExtract} outlines the \Call{findExtract}{} function.
To begin with, $\operatorname{findLS}()$ finds the longest common substring $s$ shared by b[i], which is an insertion edit, and the entire buckets.
The subscription $r$ requests to match the \textbf{r}hs of insertion b[i] to the lhs of deletion or substitution edits in buckets.
In this way, we find which sites the ``function body''--a metaphor for the extracted text--is moved from.

\begin{algorithm}[t]
\begin{algorithmic}[1]
\Function{findExtract}{i, b, buckets, P\_ref}
\State s, sites $\gets \operatorname{findLS}_r$ (b[i], buckets)
\State $P_c \gets \varnothing$
\State \Call{expandEdit}{i, b, buckets, $P_c$}
\State $c, m_c \gets $ \Call{SortAndFilter}{$P_c$}[0]
\State c.rhs $\gets$ c.rhs $\circ (s\to \magnet) $
\State $P_a \gets \varnothing $
\ForAll{j $\in$ sites}
\State \Call{expandEdit}{j, b, buckets, $P_a$}
\EndFor
\State $a, m_a \gets $\Call{SortAndFilter}{$P_a$}[0]
\State a.lhs $\gets$ a.lhs $\circ (s\to \magnet) $
\State P\_ref[$a\vdash c$] $\gets m_a + m_c$
\EndFunction
\end{algorithmic}
\caption{Build a move rule for many-to-one extractions}
\label{algo:findExtract}
\end{algorithm}

Then from Line 3 to 4, \Call{findExtract}{} makes use of \Call{expandEdit}{} defined in Algorithm~\ref{algo:substitution}
to find a substitution rule as the consequent for inserting the new ``function definition''--the ``function body'' plus an optional prefix and postfix.
From Line 7 to 10, we loop the sites where the longest substring is found and create an antecedent rule for deleting the sites.
Furthermore, in Line 6 and 12 the rewriting rule s $\to$ {\magnet} is applied onto the rhs of $c$ and the lhs of $a$ respectively,
to install a capturing device and a backreference device.
Finally, a move rule is created and its classification metrics (tp and fp) is the sum of its components (see Line 5, 11, and 13).
Apparently, if $s$ is too trivial, $P_c$ or $P_a$ is empty, then \Call{findExtract}{} terminates early.
The function \Call{findInline}{} is quite the opposite, which involves calling $\operatorname{findLS}_l$.

In terms of complexity, $\operatorname{findLS}$ finishes in $O(\frac{N}{M}M)=O(N)$.
\Call{expandEdit}{} finishes in $O(\frac{w^2N^2}{M})$ which we have shown in the previous subsection.
Then, let $d$ be the number of insertion or deletion edits ($N \ge M\ge d\ge 0$).
The time complexity of \Call{findExtract}{} becomes $O\left(N+(M-d)\frac{w^2N^2}{M}\right)$.
\Call{getPreciseMove}{} loops \Call{findExtract}{} $d$ times.
Therefore, the overall running time is $O(dw^2N^2)$.

\subsection{Applying Rules}\label{sec:apply-rules}

When string-rewriting rules and move rules are obtained, applying them to strings is a straightforward search-and-replace operation.
Nevertheless, our rules were created from tokenized strings, and thus each match must start and end at token boundaries.
Otherwise,
for instance, the string-rewriting rule \code{public} $\rightarrow$ \code{private} that changes the visibility of a method
will replace the class name \code{Republican} with \code{Reprivatean}.

{\Summer} does not expect a rule to be applied certain number of times
because the situation of the left commit and the right commit can differ.
For example, the left commit changes the indentation style from 4 spaces to 1 tab \code{\CE{s}\CE{s}\CE{s}\CE{s}} $\rightarrow$  \code{\CE{t}},
and the rule applies at 100 places in the left commit.
When the rule is used in the right revision, as this revision has already been formatted to tabs, this rule applies few times.
The disparity between application counts is not a red flag.

Given a set of bucketed string edits,
{\Summer} first tries Section 3.1, 3.2, 3.3, 3.5, 3.6 for non-overlapping move rules.
When the algorithm in Section 3.5 finds no move rule, {\Summer} executes Section 3.1, 3.2, 3.3, 3.4, 3.6 for non-overlapping string-rewriting rules.
Accordingly, {\Summer} may run twice for a complicated commit.

\section{Evaluation by a Modified ConflictBench}\label{sec:conflictBench}

A number of benchmark suites have been proposed for testing merge tools;
however, many are no longer available, such as \etal{Seibt}'s dataset~\cite{seibt2021leveraging} and \Tool{IntelliMerge}'s dataset~\cite{IntelliMerge}.
The data used to train \Tool{DeepMerge}~\cite{dinella2022deepmerge} is limited, including only resolutions achieved by rearranging lines.
Similarly, datasets that were published with refactoring-aware merge tools include only conflicts induced by code refactorings.
In contrast, \etal{Ghiotto}~\cite{ghiotto2018nature} provided a general dataset of 25,328 merge failures.
\etal{Shen} composed ConflictBench, which has 180 merge failures~\cite{shen2024conflictbench}.
We eventually chose ConflictBench because it was new and easy to set up.

The merge driver in Shen's ConflictBench aims to create merge results by giving merely 3 files and their paths to a merge tool, corresponding to the base, the left, and the right branch.
The merge driver does not evaluate merge results, which is done manually by checking semantic equivalence.
The 3 files by design were not mergeable by \code{git merge}.
This benchmark comes with the binaries of 4 merge tools, namely \Tool{AutoMerge}, \Tool{FSTMerge}, \Tool{IntelliMerge}, and \Tool{JDime};
and it installs \Tool{KDiff3} during setup.
We did not seek latest versions of these tools in order to produce fair and comparable results.
\Tool{KDiff3} is a GUI tool and all others are command line.

We introduce a fully automatic method for assessing the literal similarity of merge results.
Although solo developers focus more on semantic equivalence,
IT companies have strict requirements on code styles and readability and they may favor literal equivalence.
Moreover verbatim comparison is purely objective and automatic.
Therefore, we modified ConflictBench to call \code{git diff -{}-ignore-blank-lines -{}-ignore\allowbreak{}-all-space} to automatically compare the file synthesized by a merge tool and the one by the developer.
For Java files, we make a minimal normalization by sorting the import statements.
Put simply, \textbf{our modified ConflictBench automatically computes merge results' literal similarity by giving merely 3 files to a merge tool}.

As of the dataset in ConflictBench,
the repo \repo{mybatis-plus} contains a strange folder named \code{d:$\backslash$codeGen} which is not accepted by Windows.
This folder is outside of the tree structure of the conflicting file, so
we mirrored this repo and removed this folder without affecting the merge process.
The repo \repo{FEBS-Shiro} was no longer available and we could not find any forks.
Hence, the total number of repos was reduced to \datasetSize.
Although the main language of these repos is Java, {\totalNonJava} merge scenarios are about non-Java files,
a breakdown of which is displayed in Table~\ref{nonJavaStats}.

\begin{table}
\centering
\caption{The types of {\totalNonJava} non-java files in the dataset}\label{nonJavaStats}
\begin{tabular}{ll}
File Type         & Count 				\\\hline
pom.xml           & \nonJavaXml    		\\
txt, markdown, or adoc & \nonJavaMarkdown    \\
build.gradle      & \nonJavaGradle     	\\
properties files  & \nonJavaProperties     \\
groovy or scala   & \nonJavaLikeJava    \\
others			  & \nonJavaOthers
\end{tabular}
\end{table}

Given that four merge tools rely on abstract syntax trees,
we added another baseline tool \Tool{wiggle}~1.3, as a representative of purely textual merge approaches.
Since \Tool{KDiff} is a GUI tool,
we coded an AutoHotKey script to click the merge button and close dialogs which interrupt the automatic process.
\Tool{AutoMerge} extended \Tool{JDime}, so we skipped the latter in our experiments.

Besides, we made additional improvements to Shen's merge driver while preserving its behavior.
The original merge driver prepares the testing input by copying a repo folder and deleting everything but the single conflicting file,
which easily leads to file locking issues and breaks the integrity of a test run.
Our new version utilizes the sparse checkout feature of Git so that only the required file is materialized, greatly boosting the robustness of our evaluation.

\begin{table*}
\centering
\caption{Evaluation results divided by the Java category and the non-Java category.
Each category shows the number of solutions literally (lit.) and semantically (sem.) matching developers',
and their accuracy.}
\label{table:evaluation-results}
\begin{tabular}{l|l|llllll}
&                & Summer & KDiff3 & wiggle & FSTMerge & IntelliMerge & AutoMerge \\ \hline
	\multirowcell{5}{{\totalJava}\\ Java\\ cases}
& Lit.\ matches  & \SummerMatchesJava     & 12     & 4      & 5        & 5            & 0         \\
& Sem.\ matches  & -      & 15     & -      & 24       & 29           & 49        \\
& Lit.\ Accuracy & \SummerAccuracyJava   & 11\%   & 3.8\%   & 4.7\%     & 4.7\%         & 0      \\
& Sem.\ Accuracy & -      & 14\%   & -      & 23\%     & 27\%         & 46\%      \\
& Merge Accuracy & \SummerAccuracyJava   & 11\%   & 3.8\%   & 23\%     & 27\%         & 46\%      \\ \hline
\multirowcell{5}{{\totalNonJava}\\ non-Java\\ cases}
& Lit.\ matches  & \SummerMatchesNJ     & 11     & 1      & 4        &              &           \\
& Sem.\ matches  & -      & 14     & -      & 9        &              &           \\
& Lit.\ Accuracy & \SummerAccuracyNJ   & 15\%   & 1.4\%   & 5.5\%     &              &           \\
& Sem.\ Accuracy & -      & 19\%   & -      & 12\%     &              &           \\
& Merge Accuracy & \SummerAccuracyNJ   & 15\%   & 1.4\%   & 12\%     &              &           \\ \hline
Total
& Merge Accuracy & \SummerAccuracyOverall  & 13\%   & 2.8\%   & 18\%     &          &
\end{tabular}
\end{table*}

The evaluation results are in Table~\ref{table:evaluation-results}.
The first 3 tools, namely \Summer, \Tool{KDiff}, and \Tool{wiggle}, are universal merge tools.
\Tool{FSTMerge} is not universal, but it can merge Java, XML files, and a number of other formats.
\Tool{IntelliMerge} and \Tool{AutoMerge} were designed to merge Java files, so we left their values empty in the  non-Java cases.
The literal matches were automatically reported by \code{git diff}.
The values in the row semantic matches were copied from Shen's report.
We define \textbf{merge accuracy} as literal match accuracy for a textual merge tool,
and as semantic match accuracy for an AST tool.

To summarize,
in the Java category,
the {\SummerMatchesJava} matches made by {\Summer} are the highest in the literal matching track.
If we compare merge accuracy,
{\Summer} ranks second losing to \Tool{AutoMerge}.
In the non-Java category, {\Summer} achieved the highest accuracy across all metrics.
For corporations that have strict rules regarding code styles and readability,
we recommend {\Summer} because it observes the original, local styling and syntax,
it does not delete comments and copyright headers,
and its merge output is ready to commit.

\begin{table}
\caption{{\Summer}'s unique solutions to each other tool, and other tools' unique solutions to {\Summer}.
We analyze the repos of which names have stars.}
\label{table:venn}
\begin{subcaptionblock}{\columnwidth}
\centering
\caption{the Java category}
\begin{tabular}{l|l|l|l}
Summer-only & Both   & Other-only &              \\\cline{1-3}
25          & 9      & 3          & KDiff3       \\\cline{1-3}
34          & 0      & 4          & Wiggle       \\\cline{1-3}
33          & server* & 4          & FSTMerge     \\\cline{1-3}
31          & 3      & \makecell[l]{thumbnailator*\\JCTools}          & IntelliMerge \\\cline{1-3}
11          & 23      & 26          & \makecell[l]{AutoMerge's \\semantic matches}
\end{tabular}
\end{subcaptionblock}
\\\vspace{1em}
\begin{subcaptionblock}{\columnwidth}
\centering
\caption{the non-Java category}
\begin{tabular}{l|l|l|l}
Summer-only & Both & Other-only &          \\\cline{1-3}
23          & 7    & 4          & KDiff3   \\\cline{1-3}
29          & aeron    & 0          & Wiggle   \\\cline{1-3}
27          & 3    & jmonkeyengin*          & FSTMerge
\end{tabular}
\end{subcaptionblock}
\end{table}

In the subsequent subsections,
we first analyze a few most difficult cases for all merge tools.
Then we use Table~\ref{table:venn} to guide our analysis.
This table is venn diagrams showing intersections of {\Summer}'s solutions and each other tool's solutions,
and solutions unique to {\Summer} and to the other.
When the size of a set is less than 3, we show the repo names instead of a number.
As a result, we analyze the repos \repo{server}, \repo{thumbnailator}, and \repo{jmonkeyengin},
which have stars in Table~\ref{table:venn}.
Finally, we discuss the disparity between \Tool{AutoMerge}'s literal matches and semantic matches by checking additional repos.

\subsection{Java Scenarios}

The repo \repo{junit4} is one of the 59 most difficult cases, which no tool can merge correctly.
In \repo{junit4}, the left branch increased a version number; the right branch removed the snapshot tag from the version number;
and the merge commit added back the snapshot tag.
This case is particularly challenging in that the correct resolution requires a solid understanding of the meaning of the snapshot tag.
Another most difficult case is \repo{halo}, where the left branch added a comment warning about a null pointer exception, and the right branch fixed the null pointer exception.
The developer only adopted the right branch.
A large language model should be able to merge the local region correctly.

\begin{figure}
\setlength{\abovecaptionskip}{0pt}
\captionof{lstlisting}{For \repo{thumbnailator}, {\Summer} (on the left) added an extra \code{since} tag comparing to developer.
\Tool{FSTMerge} and \Tool{Auto\-Merge} (on the right) wrote in a style differing to developer}
\label{FSTMerge-summer-thumbnailator}
\begin{minipage}[t]{0.45\columnwidth}
\begin{lstlisting}[language=diff, tabsize=8]
  * 
  * @author coobird
  * @since	0.3.4
- * @since 0.3.4
  *
  */
 public interface Size
\end{lstlisting}
\end{minipage}
\hfill
\begin{minipage}[t]{0.45\columnwidth}
\begin{lstlisting}[language=diff]
  * @since      0.3.4
  *
  */
-public  interface  Size {
-
+public interface Size
+{
        /**
         * Calculates the size

\end{lstlisting}
\end{minipage}
\end{figure}

The repo \repo{thumbnailator} is one of the two cases that \Tool{IntelliMerge} could handle but {\Summer} could not, as shown in Table~\ref{table:venn}.
In \repo{thumbnailator}, its left branch adds a \code{@since} tag followed by a space, and its right branch adds a \code{@since} tag followed by a tab.
As {\Summer} does not know semantics of the space and the tab character, it adds both to the merge result, shown in Listing~\ref{FSTMerge-summer-thumbnailator}.
On the contrary, the developer picked merely the tab version.
\Tool{FSTMerge} and \Tool{AutoMerge} resolved the conflicts logically correctly, but did not respect the original coding style, shown in the same listing.
\Tool{IntelliMerge} successfully merged this case. \Tool{KDiff} and \Tool{wiggle} both gave up.

\begin{lstlisting}[language=diff, float, label={IntelliMerge-server}, caption={\Tool{IntelliMerge} produced duplicated \code{throws} expressions for \repo{server}}]
     boolean received;
     Server server;

-    protected void startServer() throws IOException throws IOException {
+    protected void startServer() throws IOException {
         server = new Server();
         server.startServer();
     }

     @Before
-    public void setUp() throws IOException throws IOException {
+    public void setUp() throws IOException {
         startServer();
     }
\end{lstlisting}

Although \Tool{IntelliMerge} claims to be an AST tool, it does not always produce syntactically correct output.
A case in point is the repo \repo{server}.
Listing~\ref{IntelliMerge-server} shows the merge result by \Tool{IntelliMerge} for this repo,
where the duplicated \code{throws} expressions are not legal.
Only {\Summer} and \Tool{FSTMerge} passed this case.

\begin{lstlisting}[language=diff, float, label={AutoMerge-SimianArmy}, caption={\Tool{AutoMerge} failed to rename a variable in the merge result for \repo{SimianArmy}}]
-  @Override
-  public void doMonkeyBusiness() {
+  /** {@inheritDoc} */
+  @Override public void doMonkeyBusiness() {
     allChaosTypes = Lists.newArrayList();
     allChaosTypes.add(new ShutdownInstanceChaosType(cfg));
-    enabledChaosTypes.add(new DetachVolumesChaosType(cfg));
+    allChaosTypes.add(new DetachVolumesChaosType(cfg));
   }
\end{lstlisting}

The repo \repo{SimianArmy} is one of the 11 cases in Table~\ref{table:venn} which
{\Summer} can literally match and \Tool{AutoMerge} cannot semantically match.
In \repo{SimianArmy}, the left branch adds a new line
\code{enabled\-ChaosTypes.add(new DetachVolumesChaosType(cfg));}
and the right branch renames all \code{enabledChaosTypes} to \code{allChaosTypes}.
Listing~\ref{AutoMerge-SimianArmy} shows that in addition to formatting and comment issues,
\Tool{AutoMerge} added the new line to the merge result, but failed to rename it to \code{allChaosTypes}.
{\Summer} passed this case. All other tools failed.

\begin{lstlisting}[language=diff, float, label={AutoMerge-RxJava},
caption={\Tool{AutoMerge} printed a full syntax tree without omitting optional elements for \repo{RxJava}}]
+/**
+ * Copyright 2014 Netflix, Inc.
+ * 
+ * Licensed under the Apache License, Version 2.0 (the "License"); you may not use this file except in
+ ...
+ */
 package rx;
 
...

-  @SuppressWarnings(value = { "unchecked" }) public final static <T extends java.lang.Object> Observable<T> concat(...) {
+    @SuppressWarnings("unchecked")
+    public final static <T> Observable<T> concat(...) {
         return create(OperationConcat.concat(t1, t2, t3));
     }
...
 
-  private static class ThrowObservable<T extends java.lang.Object> extends Observable<T> {
+    /**
+     * An Observable that invokes {@link Observer#onError onError} ... 
+     * @param <T> the type of item ...
+     */
+    private static class ThrowObservable<T> extends Observable<T> {
	     public ThrowObservable(final Throwable exception) {
\end{lstlisting}

\Tool{AutoMerge}'s solutions deviate from developers' not only in comments and line breaks, but also in optional elements in the Java language.
Shen classified \Tool{AutoMerge}'s solution in Listing~\ref{AutoMerge-RxJava} as semantically equivalent,
but the difference is huge in a literal sense.
For starters, \Tool{AutoMerge} did not preserve the copyright header and javadoc because they are comments.
In an enterprise context, this merge result may violate licensing requirements.
Secondly, \Tool{AutoMerge} printed optional nodes in the syntax tree, such as
\code{value = \{ "unchecked" \}} versus simply \code{"unchecked"},
and \code{T extends java.lang.Object} versus simply \code{T}.
Only \Tool{KDiff} and {\Summer} passed this test.

\subsection{Non-Java Scenarios}

There are 38 most difficult cases in the non-Java scenarios where no tool can merge correctly.
One of them is repo \repo{redisson}. The correct value in an XML node is \code{3.4.2-\linebreak[1]SNAPSHOT}.
The value {\Summer} gave is \code{3.4.3-SNAPSHOT}, while \Tool{wiggle} gave \code{2.9.3-SNAPSHOT}.
\Tool{FSTMerge} formatted the XML file in its own way, but the value of the version tag was correct.
\Tool{KDiff} refused to merge.
Another case is \repo{reactive-streams-jvm}.
The file in conflict is \code{Copyright\-Waivers.\linebreak[1]txt}, which hosts a tabular structure.
Shown in Listing~\ref{fstmerge-reactive}, the output of \Tool{FSTMerge} did not have the rows ouertani, 2m, and ldaley.
{\Summer}, on the other hand, added all necessary rows but did not order them to the developer's liking.
\Tool{KDiff} and \Tool{wiggle} refused to merge.

\begin{figure}
\begin{lstlisting}[language=diff, label={fstmerge-reactive},
caption={\Tool{FSTMerge} failed to add certain rows in \repo{reactive-streams-jvm}}]
 savulchik      | Stanislav Savulchik, s.savulchik@gmail.com
 ktoso          | Konrad Malawski, konrad@project13.pl, Typesafe 
+ouertani       | Slim Ouertani, ouertani@gmail.com
+2m             | Martynas Mickevicius, mmartynas@gmail.com, Typesafe 
+ldaley         | Luke Daley, luke.daley@gradleware.com, Gradleware 
 colinrgodsey   | Colin Godsey, crgodsey@gmail.com, MediaMath Inc.
\end{lstlisting}
\begin{lstlisting}[language=diff, label={summer-reactive},
caption={{\Summer} added a row to a wrong place in \repo{reactive-streams-jvm}}]
 savulchik      | Stanislav Savulchik, s.savulchik@gmail.com
 ktoso          | Konrad Malawski, konrad@project13.pl, Typesafe 
-colinrgodsey   | Colin Godsey, crgodsey@gmail.com, MediaMath Inc.
 ouertani       | Slim Ouertani, ouertani@gmail.com
 2m             | Martynas Mickevicius, mmartynas@gmail.com, Typesafe 
 ldaley         | Luke Daley, luke.daley@gradleware.com, Gradleware 
+colinrgodsey   | Colin Godsey, crgodsey@gmail.com, MediaMath Inc.
\end{lstlisting}
\end{figure}

The left and right branches of the repo \repo{jmonkeyengine} made similar changes,
both having
\code{-J-XX:PermSize=128m} $\Rightarrow$
\code{-J-XX$\backslash$:PermSize$\backslash$=\linebreak[1]128m}.
The merge result of {\Summer} is \code{-J-XX$\backslash\backslash$:PermSize$\backslash\backslash$=\linebreak[1]128m}
because {\Summer} incorrectly applied the rule \code{:} $\rightarrow$ \code{$\backslash$:} twice
and \code{=} $\rightarrow$ \code{$\backslash$=} twice.
This case reveals that our algorithm can make mistakes with rules that insert characters.

\section{Threats to Validity}\label{sec:threats-to-validity}

Based on the evaluation of {\datasetSize} Java and non-Java cases, we conclude that
{\Summer} outperforms other universal tools, though specialized AST tools achieve higher semantic accuracy.
However, there are threats to the validity in our study.

\subsection{Construct Validity}

Our evaluation measures ``merge accuracy'', i.e., literal match accuracy for a textual merge tool and semantic match accuracy for an AST tool.
The literal match accuracy is purely objective and the semantic match accuracy is subjective.
For example, it is subjective whether files with and without a copyright header are semantically equivalent.
We relaxed the literal matching criteria by ignoring blank lines and spaces, and sorting Java's \code{import} statements.

\subsection{Internal Validity}
Bugs in the implementation of the described algorithm threaten the internal validity of our claim,
but we added 40+ automated tests and numerous assertions to ensure the correctness of {\Summer}.

The validity of Shen's original ConflictBench has been established in \cite{shen2024conflictbench}.
To avoid introducing new threats, we preserved the original behavior of the benchmark driver and kept our improvements minimal,
while enhancing  error detection and reducing crashes due to file locking.
Removing the repo \repo{FEBS-Shiro} may slightly affect the dataset balance.
However, we did not add a new repo because it would introduce a new bias.
We fixed a directory in \repo{mybatis-plus} that was invisible to a merge tool,
so it had no impact on validity.
Overall, our changes were necessary for a consistent and fair platform
to establish experimental reliability.

\subsection{External Validity}

The evaluation was mainly done on the Java programming language with a variety of other data formats.
Consequently, the accuracy numbers may not generalize well to documents in other languages or formats.

{\Summer} currently ignores changes in file encoding and line endings.
These kinds of changes exist in the real world, but they are considered infrequent.
Also, ConflictBench does not test multiple conflicting files.
If one branch deletes file A and modifies file B, and the other branch modifies file A and deletes file B,
as discussed in Section~\ref{sec:merge-direction},
a suitable merge direction cannot be determined and this reduces {\Summer}'s merge accuracy.

Finally, ConflictBench only provides the conflicting file to each merge tool,
but {\Summer} actually can read all changed files in the left and right branches to determine the best merge.
Receiving limited information, {\Summer}'s accuracy is a conservative estimate.
The same applies to other tools.
ConflictBench does not send styling guides to AST tools
and thus they made formatting mistakes seen in Listings~\ref{FSTMerge-summer-thumbnailator}, \ref{AutoMerge-RxJava}, and \ref{fstmerge-reactive}.
To mitigate this threat,
the merge accuracy is defined as semantic match accuracy for an AST tool relying on Shen's manually evaluated accuracy numbers.

\section{Conclusions}

A universal merge tool offers two key advantages in software engineering.
Firstly, a single universal merge tool has low maintenance burdens on both developers and users.
Its developers do not have to update the tool when a specific programming language introduces new syntax.
Users avoid installing multiple language-specific tools, each requiring security vetting in enterprise environments.
Additionally, multiple tools create compatibility challenges, as their dependencies may conflict.
Secondly, advances in universal merge algorithms boost performance of high-level syntactic or semantic merge tools
because high-level tools rely on string merge algorithms for the content of terminal nodes in a syntax tree, for instance the literal value of a string node.

Our proposal {\Summer}, as a universal textual merge algorithm and tool,
observes original coding styles and deliberate syntax choices,
preserves comments and copyright headers,
and its merge output is ready to commit.

For Java files, without relying on the knowledge about syntax or code refactoring,
{\Summer} matched {\SummerAccuracyJava} of developers' merge resolutions character-to-character.
The accuracy for XML, ReadMe, and other files is {\SummerAccuracyNJ},
and overall {\SummerAccuracyOverall}.
The percentages are higher than other well-developed merge tools,
demonstrating the effectiveness of the proposed algorithm.

In the future, we plan to integrate our algorithm with the Darcs version control system.
Next, we need to test {\Summer} on a larger dataset, for example Ghiotto's~\cite{ghiotto2018nature}.

\subsubsection*{Data Availability}
The source code of {\Summer} is available at \cite{SourceSummer}.
The source code of the improved ConflictBench is at \cite{SourceConflictBench}.

\subsubsection*{Acknowledgements}
The research was supported by the European Union under the project
\textsf{ROBOPROX} (reg.~no.~CZ.02.01.01/00/22\_008/\linebreak[1]0004590).
This article is part of the \textsf{RICAIP} project that has received funding
from the European Union's Horizon~2020 research and innovation programme under
grant agreement No~857306.

\ifdefined\CheckPageLength\else%
\bibliographystyle{ACM-Reference-Format}
\bibliography{bib.bib}\fi
\end{document}